\documentclass[fleqn,12pt]{wlscirep}
\usepackage[T1]{fontenc}
\usepackage{bm}
\usepackage{xcolor}
\usepackage{float}
\DeclareUnicodeCharacter{2212}{-}

\title{Optically-triggered strain-driven N\'{e}el vector manipulation in a metallic antiferromagnet}

\author[1,2]{Vladimir Grigorev}
\author[1,2,3]{Mariia Filianina}
\author[1]{Yaryna Lytvynenko}
\author[1]{Sergei Sobolev}
\author[1]{Amrit R. Pokharel}
\author[4]{Alexey Sapozhnik}
\author[5]{Armin Kleibert}
\author[6]{Stanislav Yu. Bodnar}
\author[7]{Petr Grigorev}
\author[8]{Yurii Skourski}
\author[1,2]{Mathias Kl\"aui}
\author[1,2]{Hans-Joachim Elmers}
\author[1]{Martin Jourdan}
\author[1,2]{Jure Demsar}

\affil[1]{Institute of Physics, Johannes Gutenberg University, Staudingerweg 7, 55128 Mainz, Germany}
\affil[2]{Graduate School of Excellence Materials Science in Mainz, Staudingerweg 9, 55128 Mainz, Germany}
\affil[3]{Department of Physics, AlbaNova University Center, Stockholm University, S-106 91 Stockholm, Sweden}
\affil[4]{Institute of Physics, LUMES, École Polytechnique Fédérale de Lausanne (EPFL), Lausanne,
Switzerland}
\affil[5]{Paul Scherrer Institute, Swiss Light Source, 5232 Villigen PSI, Switzerland}
\affil[6]{Walter Schottky Institute and Physics Department, Technical University Munich, 85748 Garching, Germany}
\affil[7]{Aix-Marseille Universit\'{e}, CNRS, CINaM UMR 7325, Campus de Luminy, 13288 Marseille, France}
\affil[8]{Dresden High Magnetic Field Laboratory (HLD-EMFL), Helmholtz-Zentrum
Dresden-Rossendorf, 01328 Dresden, Germany}

\begin{abstract}
The absence of stray fields, their insensitivity to external magnetic fields, and ultrafast dynamics make antiferromagnets promising candidates for active elements in spintronic devices. Here, we demonstrate manipulation of the N\'{e}el vector in the metallic collinear antiferromagnet Mn$_2$Au by combining strain and femtosecond laser excitation. Applying tensile strain along either of the two in-plane easy axes and locally exciting the sample by a train of femtosecond pulses, we align the N\'{e}el vector along the direction controlled by the applied strain. The dependence on the laser fluence and strain suggests the alignment is a result of optically-triggered depinning of 90$^{\mathrm{o}}$ domain walls and their sliding in the direction of the free energy gradient, governed by the magneto-elastic coupling. The resulting, switchable, state is stable at room temperature and insensitive to magnetic fields. Such an approach may provide ways to realize robust high-density memory device with switching timescales in the picosecond range.
\end{abstract}

\begin{document}

\flushbottom
\maketitle

\thispagestyle{empty}

Antiferromagnets (AFM) have recently attracted major scientific interest
due to their prospective applications in the field of spintronics.\cite%
{spintronics_2,spintronics_4,spintronics_5,Nemec} For information storage,
they offer several advantages over ferromagnets, such as a potential for
ultrafast switching, the lack of stray fields and, thus, the robustness
against external fields. All these properties could enable high information density, with the direction of the staggered magnetization (N\'{e}el vector, 
\textbf{L}) as the information carrier. The lack of sensitivity to
external magnetic fields, however, poses challenges in terms of writing \cite%
{Wadley_Sci_2016,OleTHz,Stas_Nature} and read-out \cite%
{CuMnAs_PRL,Stas_Nature,Schreiber,dichroism,Sempa} schemes. Exploring the ways to rapidly and reliably manipulate the N\'{e}el vector direction
is at the heart of the current AFM spintronics research.\cite{spintronics_2,spintronics_4,spintronics_5,Nemec} Especially the metallic
collinear antiferromagnets CuMnAs and Mn$_{2}$Au, which enable current-induced bulk (Néel) spin-orbit torques,\cite{Zelezny} have been in the
focus of recent research. Several approaches for switching the direction of the N\'{e}el vector have been suggested, which include aligning the N\'{e}el vector by high magnetic fields,\cite{Leha_PRB,Gomonay} application of strain,\cite{Leha_PSS,Piezo}
and, most prominently, using current-driven N\'{e}el spin-orbit torques.\cite%
{Wadley_Sci_2016,OleTHz,Stas_Nature} In terms of the read-out, in addition
to X-ray magnetic linear dichroism,\cite{Leha_PRB,Piezo,Wadley_imaging} electrical read-out via the anisotropic magnetoresistance (AMR)\cite%
{Wadley_Sci_2016,Stas_Nature} and optical magnetic linear dichroism (OMLD)%
\cite{Saidl2020,dichroism} have been realized. In spite the achieved progress, an ultrafast approach for locally changing the N\'{e}el vector orientation is needed to fully exploit the potential of AFMs in spintronics.

In this work, we demonstrate a novel approach to manipulate the N\'{e}el
vector in a collinear metallic antiferromagnet Mn$_{2}$Au by combining applied strain with femtosecond laser excitation. Exciting the strained Mn$_{2}$Au film by a train of femtosecond near-infrared pulses, we align the N\'{e}el vector in the direction perpendicular to the applied tensile strain. Imaging the resulting magnetic domain structure by photoemission electron microscopy using X-ray magnetic
linear dichroism to obtain magnetic contrast (XMLD-PEEM), we demonstrate the
existence of a threshold fluence required for N\'{e}el vector alignment,
which depends on the strain. Analyzing the magnetic domain
structures before and after optical excitation, we attribute the observed
optically activated N\'{e}el vector alignment in Mn$_{2}$Au to depinning,
motion and annihilation of the 90$^{\mathrm{o}}$ domain walls. Based on the threshold fluence values and their dependence on strain, we suggest the depinning is
activated by laser induced transient heating. The demonstrated method of local optical manipulation of the N\'{e}el vector can be applied to a range of collinear AFMs with low magneto-crystalline anisotropies, enabling ultrafast writing of magnetically stored information. In combination with fast optical readout schemes,\cite{Saidl2020,dichroism} this approach provides means for studying
switching and domain wall dynamics on an ultrafast timescale.

\section*{Laser writing under tensile strain}

The metallic collinear AFM Mn$_{2}$Au has a body-centered tetragonal crystal
structure. It is an easy plane AFM, with a strong out-of-plane [001] hard
axis and a weak 4-fold in-plane magnetic anisotropy. The Mn spin orientations in
adjacent planes are anti-parallel, with \textbf{L} pointing along
one of the easy $\langle 110\rangle $ directions.\cite{Barthem,FilmsAFM}
The $c$-axis epitaxial 45 nm thin Mn$_{2}$Au films are grown on $r$-cut (1$\overline{%
1}$02) Al$_{2}$O$_{3}$ substrates, with a 13 nm Ta (001) buffer layer for
improving epitaxial growth.\cite{FilmsAFM} The $\langle 110\rangle $ directions of Mn$_{2}$Au are parallel to the substrate edges, which are along the $[010]_{s}$ and $[211]_{s}$ directions of the $r$-cut Al$_{2}$O$_{3}$ substrate (index $s$ stands for the substrate). We note, that recent optical study on similar films revealed a weak in-plane optical linear dichroism in the as-grown films, which was attributed to a weak, growth-induced, strain.\cite{dichroism} To take this underlying asymmetry into account, we label the two easy axes such that $[110]\Vert \lbrack 010]_{s}$ and $[1%
\overline{1}0]\Vert \lbrack 211]_{s}$.

\begin{figure}[pth]
\centerline{\includegraphics[width=170mm]{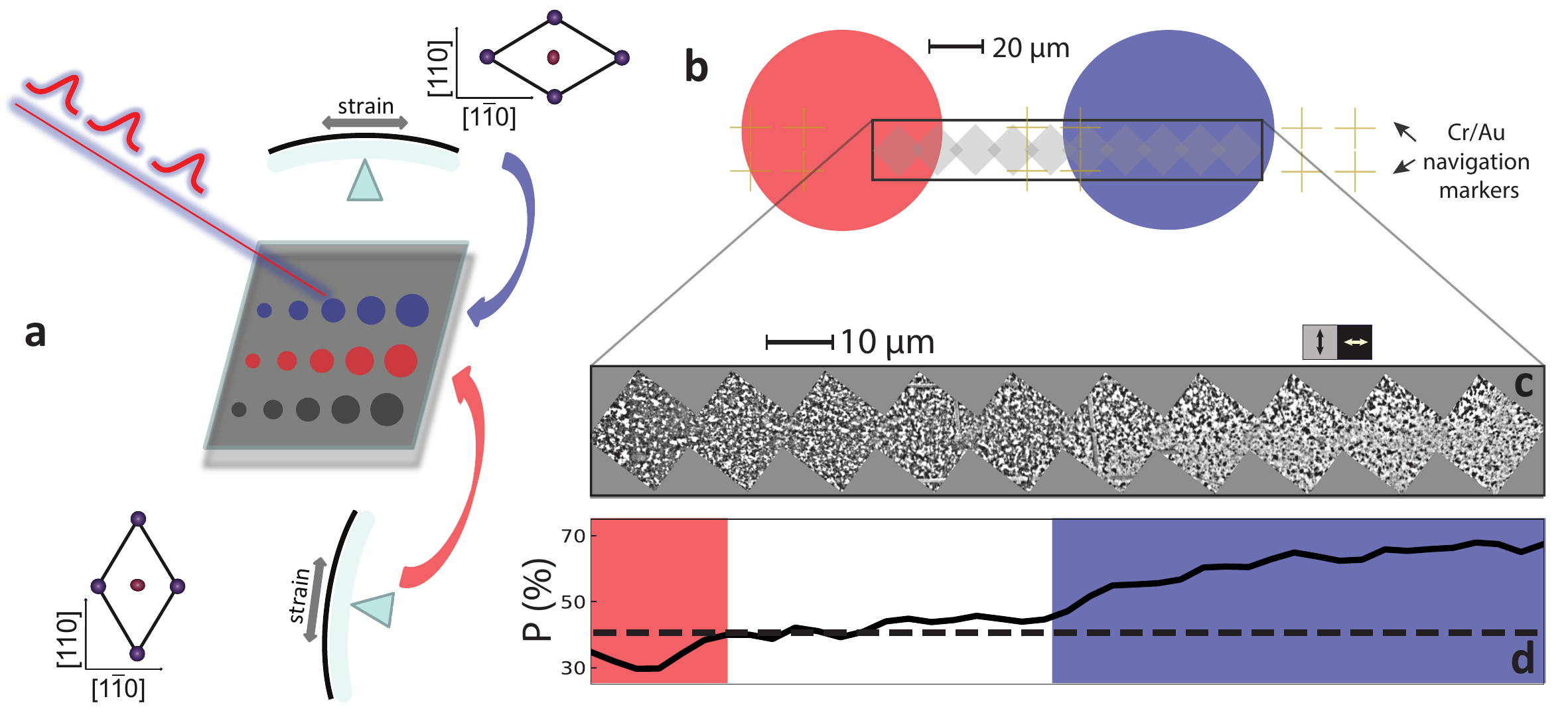}}
\caption{\textbf{Imaging the magnetic domain structure along the line connecting two areas, irradiated under orthogonal tensile strain.} \textbf{a} Schematics of the experimental approach for testing the strain-assisted light-induced N\'{e}el vector switching. The red (blue) circles represent the areas on the film irradiated under the tensile strain applied along [110] ([1$\overline{1}$0]) direction, while the black circles represent areas irradiated in the absence of external strain. The size of the circle is proportional to the laser fluence. \textbf{b} Schematic sample layout, showing the Cr/Au navigation markers. Two adjacent areas were laser-irradiated under orthogonal strain. Here, red (blue) shaded areas represent the irradiated areas with strain along [110] ([1$\overline{1}$0]) directions. 
\textbf{c} Composite XMLD-PEEM image along the line connecting the two
irradiated areas obtained by merging successive XMLD-PEEM images. Between the two irradiated areas the sample is in the as-grown state. In the XMLD-PEEM images, the N\'{e}el vector in the dark/black domains is
parallel to the [1$\overline{1}$0] direction while in the bright/white domains it is
parallel to the [110] direction. The relation between the XMLD contrast and the direction of the N\'{e}el vector is based on earlier XMLD studies on aligned samples.\cite{Leha_PSS,Leha_PRB,dichroism,Sempa} \textbf{d} Corresponding lateral evolution of the area fraction of domains with the N\'{e}el vector parallel to the [110] direction, $P$. Here, $P$ is calculated by averaging the XMLD contrast along vertical stripes of $\approx 3.3$~$\mu$m width. The black dashed line presents the value of $P$ obtained on the as-grown sample, extracted by analyzing XMLD-PEEM images of the non-irradiated sample areas.}
\label{Figure1}
\end{figure}

The experimental approach to manipulate the N\'{e}el vector by combining tensile strain and femtosecond laser pulse excitation is sketched in Fig. \ref{Figure1} a,b (see Methods for details). Fifteen areas on the film are marked by a set of 30 nm thick Cr/Au markers fabricated on the top of the film using electron beam lithography. The areas are irradiated by a train of 60 fs laser pulses at 800 nm with excitation fluence in different marked areas varied between $\approx 3 $ and $\approx 11$ mJ/cm$^{2}$. The irradiated areas are 60-70 $\mu$m in diameter, as indicated by the colored circles in Fig.~\ref{Figure1}b. Five areas are irradiated with the tensile strain along the $[110]$ direction, five areas with the tensile strain along the $[1\overline{1}0]$ direction and five areas without applying external strain. The value of the tensile strain is $\varepsilon =6\pm 1\times 10^{-4}$, as measured by a Si strain gauge attached to the substrate.\cite{Leha_PSS}

After laser irradiation under strain, we image the domain structure within all illuminated areas of the film using XMLD-PEEM.\cite{XPEEM} For reference, we also recorded a number of images of non-illuminated areas across the sample. To quantitatively analyze the magnetic domain structure, we binarize the magnetic contrast images using an Otsu-threshold method\cite{Otsu} and calculate the area fractions of the two types of magnetic domains. The same procedure is applied to every image.

Figure~\ref{Figure1}c presents a sequence of successive XMLD-PEEM images
acquired along the line connecting two adjacent marked areas, irradiated with fluence $F = 8.9$ mJ/cm$^{2}$ while the tensile strain was applied along the two 
orthogonal easy directions. The resulting deformations of the crystal
lattice are schematically illustrated in Fig.~\ref{Figure1}a. On the left
side, where the tensile strain was applied along the [110] direction, the
domain structure exhibits preferentially domains with $\textbf{L}
\parallel \lbrack 1\overline{1}0]$ (dark), while on the right side, where the tensile strain was applied along the [1$\overline{1}$0] direction, the majority
of domains are with $\textbf{L} \parallel \lbrack 110]$ (bright).\cite{Leha_PSS,Leha_PRB,dichroism,Sempa} Between
the two laser-irradiated areas, the domain structure is the same as in reference areas far away from the irradiated areas.

We quantify the state of magnetic alignment by $P$, the area fraction of domains with the N\'{e}el vector parallel to the [110] direction. Figure \ref{Figure1}d presents the variation of $P$ along the line connecting the two areas irradiated under orthogonal applied tensile strain. It follows that, under laser irradiation, the N\'{e}el vector aligns perpendicular to the direction of the applied tensile strain. Between the two irradiated areas we observe an uneven fraction of dark and bright domains with $P\approx 0.4$; the same value of $P$ is found across the non-irradiated (as-grown) areas of the sample.

\section*{The dependence of light-induced N\'{e}el vector alignment on excitation fluence}

To gain insight into the origin of the light-induced N\'{e}el vector
switching, we analyze the fluence dependence of $P$ in the 
range of fluences between 3.5 and 10.6 mJ/cm$^{2}$.

\begin{figure}[pth]
\centerline{\includegraphics[width=170mm]{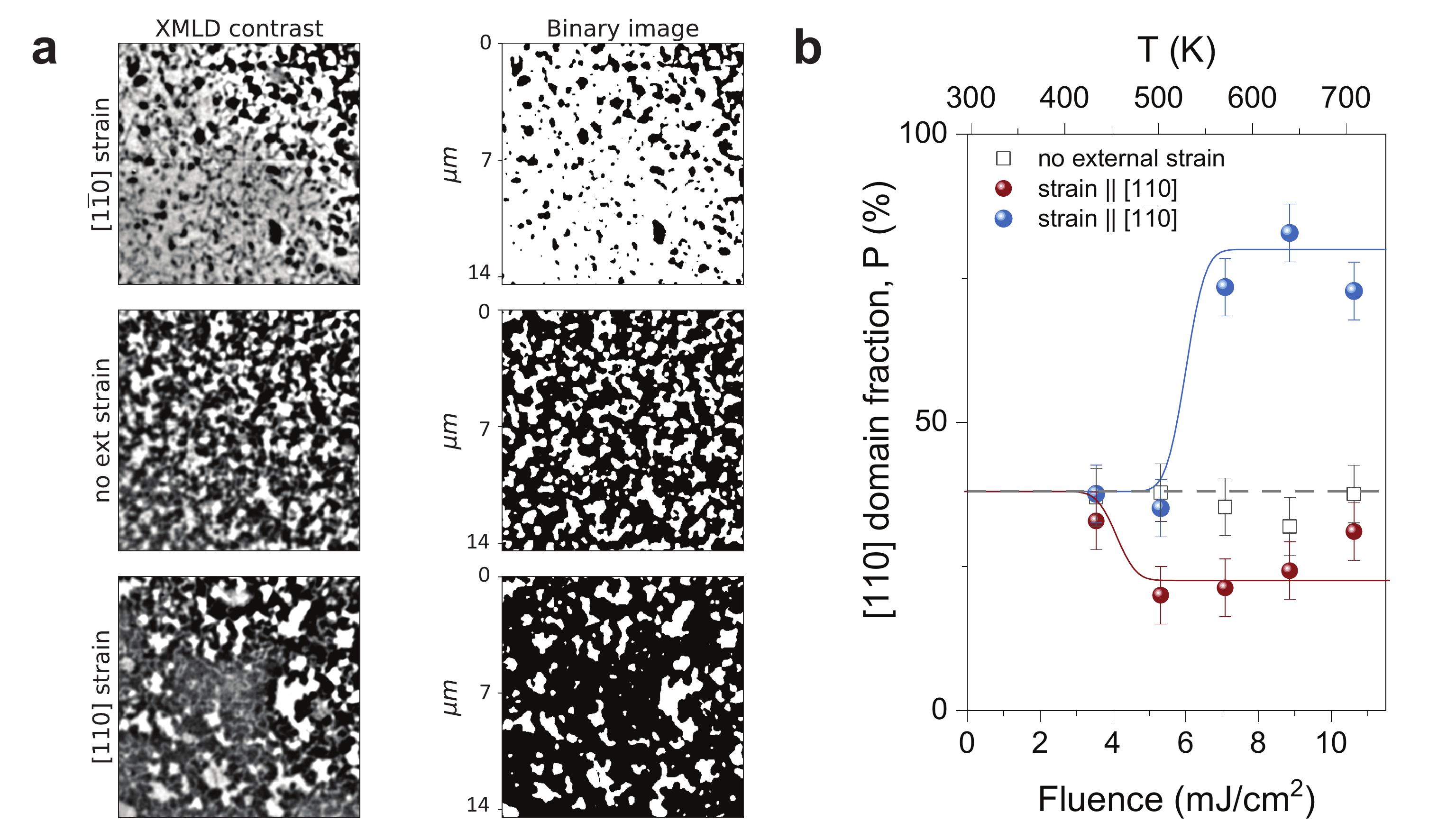}}
\caption{\textbf{Dependence of the degree of magnetic alignment on excitation fluence and strain.} \textbf{a} Left panels: XMLD contrast of areas irradiated by the near-infrared pulse trains under different external strain directions. The right panels are the corresponding binary images obtained by applying an Otsu threshold method to the images in the left panels. Top panels: strain is applied along [1$\overline{1}$0] with $F = 8.9$ mJ/cm$^{2}$; middle panels: $F = 7$ mJ/cm$^{2}$ in the absence of applied strain; bottom panels: strain applied along [110] with $F = 7$ mJ/cm$^{2}$. The binary images are used to calculate $P$ over $14\times14$~$\mu$m$^{2}$ areas. \textbf{b} Dependence of $P$ on irradiation fluence for the three strain configurations. The error-bars correspond to the variation of $P$ in the as-grown sample, obtained from XMLD-PEEM images recorded on non-irradiated areas across the film. The solid blue and red lines are guides to the eye, representing $P(F)$ for the two orthogonal directions of the applied strain, while the gray dashed line corresponds to $P$ obtained on the as-grown sample. The top horizontal axis presents the calculated peak temperatures of the irradiated volume (see Supplementary Information).}
\label{Figure2}
\end{figure}

Figure \ref{Figure2} presents the dependence of the degree of alignment
on the irradiation fluence, $F$, for all three strain configurations.
Figure \ref{Figure2}a depicts the magnetic contrast images (left column)
for sample areas irradiated under different external strain directions. From the
corresponding binary images (right column) we obtain $P$ across $14\times 14$~$\mu $m$^{2}$ imaged areas.

Figure \ref{Figure2}b displays $P(F)$ for different external strain configurations.
At $F = 3.5$ mJ/cm$^{2}$ we obtain $P\approx 0.4$
independent on the applied strain, matching the value recorded on
non-irradiated areas. The variation of $P$ at $F = 3.5$ mJ/cm$^{2}$ is similar to variation of $P$ in reference, non-irradiated, areas across the film and is governed by sample inhomogeneities. This variation can be considered as a measure of the experimental uncertainty of the extracted $P$ (error bars in Fig. \ref{Figure2}b).

The value $P\approx 0.4$ obtained for the as-grown film implies a partial N\'{e}el vector alignment. This observation is consistent with earlier imaging studies,\cite{Sempa,dichroism} and may be a result of the growth-induced strain.\cite{dichroism} Indeed, as discussed earlier,\cite{BodnarCurrentImag,dichroism} the $c$-axis of Mn$_{2}$Au films grown on a $r$-cut Al$_{2}$O$_{3}$ substrate is tilted by 2-3$^{o}$ towards the $[010]_{s}$ direction.\cite{dichroism} While no underlying strain could be resolved in the diffraction experiments,\cite{dichroism} the tilted growth and the corresponding breaking of the four-fold symmetry may indicate a weak growth-induced strain being responsible for a larger volume fraction of domains with the N\'{e}el vector perpendicular to the direction
of the $c$-axis tilt.\cite{dichroism,Sempa}

Fig. \ref{Figure2}b shows, that for both directions of the applied tensile strain there exists a threshold fluence, above which a N\'{e}el vector aligned state is realized. Here, the maximum degree of alignment, with $P$ either $\approx 0.8$ or $\approx 0.2$, is comparable to the values obtained in Mn$_{2}$Au films, where the N\'{e}el vector has been aligned in 60 T pulsed magnetic fields.\cite{Leha_PSS,Leha_PRB,dichroism}
The threshold fluence appears to be different for the two orthogonal directions of the applied strain. For the case of tensile strain applied along the $[110]$ direction, the switched state is observed already at 5.3 mJ/cm$^{2}$. On the other hand, for tensile strain applied along $[1\overline{1}0]$ the N\'{e}el vector alignment is realized at $F \gtrsim 7$ mJ/cm$^{2}$. Such a difference can be linked to the existing, growth-induced, strain in the as-grown film, which needs to be added to or subtracted from the external applied strain. Our results indicate, that the growth-induced strain could either be tensile, along the $[110]$ direction (the direction parallel to the tilt of the $c$-axis and $[010]_{s}$), or compressive, along the $[1\overline{1}0]$ direction (the direction parallel to $[211]_{s}$ and perpendicular to the direction of the $c$-axis tilt).

Finally, when irradiating the sample in the absence of external strain, $P$ did not change within the range of excitation fluences used here.

\section*{Proposed mechanism of light-induced N\'{e}el vector alignment in Mn$_{2}$Au}

In experiments, where switching of the N\'{e}el vector in Mn$_{2}$Au was
realized by applying microsecond/millisecond current pulses, the reported current-induced sample heating was significant.\cite{Stas_Nature,meinert2018,BodnarCurrentImag}  E.g., in Ref.\citenum{Stas_Nature}, a temperature increase near the switching threshold  was estimated to be 300 K. Discussing current-driven switching in granular Mn$_{2}$Au thin films, Meinert et al. proposed a phenomenological model of N\'{e}el spin-orbit torque (NSOT) - driven domain switching, assisted by sample heating.\cite{meinert2018} There, the switching was considered to be a coherent process, i.e., individual domains (the grain size with $\approx 20-30$ nm lateral dimension) were considered to switch under the NSOT. Such a switching process, however, requires overcoming an energy barrier $E_{B}\approx K_{4||}V_{g}$,\cite{meinert2018} where $K_{4||}$ is the density of the in-plane bi-axial magneto-crystalline anisotropy energy and $V_{g}$ is the domain volume. Considering a thermally activated process, with the NSOT-driven switching rate $\tau ^{-1}=\tau _{0}^{-1}\exp\left( -E_{B}/k_{B}T\right)$, a qualitative agreement between their experimental results and the Monte Carlo simulations was obtained, assuming an attempt rate $\tau _{0}^{-1}$ given by the characteristic frequency of antiferromagnets (10$^{12}$ s$^{-1}$) and $E_{B}\approx 1.5$ eV.\cite{meinert2018} With $K_{4||} \approx 1.8$~$ \mu $eV per formula unit,\cite{Sempa} such values of $E_{B}$ are expected for domains with length-scales of $\approx 20-30$ nm, as reported for Mn$_{2}$Au films grown on ZrN-buffered MgO by Meinert et al.\cite{meinert2018} However, considering the domains in our samples have a typical lateral size of 1~$\mu $m, the corresponding $E_{B}$ would be in the range of 10$^{3}$ eV. Thus, such a coherent switching of individual domains is unlikely in our epitaxial films. 

We propose the light-driven alignment of N\'{e}el vector in strained Mn$_{2}$Au films to be a result of an activated domain wall motion. We first estimate the magneto-elastic coupling constant $B_{me}$, which results in a difference in the free energy densities in orthogonally aligned magnetic domains. To this end, we performed DFT calculations of $B_{me}$ in Mn$_{2}$Au (see Supplementary Information). We obtain $B_{me}\approx 0.9$~meV per formula unit, which corresponds to a reduction in the free energy density for domains with N\'{e}el vector perpendicular to the direction of the applied tensile strain by $\approx 0.6$~$\mu $eV per formula unit for the applied strain $\varepsilon =6\times 10^{-4}$. This value is comparable to the in-plane magneto-crystalline anisotropy, $K_{4||}\approx 1.8 $~$\mu $eV per formula unit,\cite{Sempa} thus sufficient for N\'{e}el vector
alignment. Indeed, an earlier XMLD spectroscopy study on similar films did demonstrate partial N\'{e}el vector alignment under similar values of the applied external strain.\cite{Leha_PSS} However, after the applied strain was released, the original state was recovered, with no permanent N\'{e}el vector alignment.\cite{Leha_PSS} In the absence of pinning of domain walls (DWs) the applied strain should force the motion of 90$^{\mathrm{o}}$ DWs (DWs between domains with orthogonal \textbf{L}) in the direction of the free energy gradient, resulting in only two types of domains with anti-parallel \textbf{L}, separated by the 180$^{\mathrm{o}}$ DWs. While it is pinning of the 90$^{\mathrm{o}}$ DWs that prevents a stable N\'{e}el vector-aligned state to be realized using mechanical strain alone,\cite{Leha_PSS} it is likely the laser-induced depinning of DWs that is governing the strain-driven laser-assisted N\'{e}el vector-alignment.

Assuming laser-assisted depinning of DWs being the mechanism of the N\'{e}el vector alignment, we need to consider the effect of photoexcitation with a train of femtosecond optical pulses. We first note that the continuous (average) laser heating is estimated to be $<6$ K at the highest excitation densities used (see Supplementary Information) and can be neglected. Next, we estimate the induced changes in charge, spin and lattice subsystems following excitation with an ultrafast optical pulse. For simplicity, we consider the two-temperature model, that qualitatively accounts for the evolution of the electronic and lattice temperatures in metals on the (sub-)picosecond timescales.\cite{Obergfell} Here, the electron-electron scattering results in carrier thermalization on the sub-picosecond timescale. The estimated electron temperatures for the two threshold fluences are 4600 K and 5400 K (see Supplementary Information), substantially exceeding the N\'{e}el temperature of Mn$_{2}$Au ($\approx 1500$ K \cite{Barthem}). This is followed by thermalization with the lattice on a timescale of a few picoseconds and a subsequent heat diffusion into the substrate on a characteristic timescale of hundred picoseconds (see Supplementary Information). The calculated peak temperatures of the coupled electron-spin-lattice system, determined by the absorbed energy density and the total specific heat, are shown on the top axis of Fig. \ref{Figure2}b. The temperature values are comparable to those reported in the current-driven switching experiments.\cite{Stas_Nature,meinert2018,BodnarCurrentImag}

To shed light on the proposed mechanism, we performed complementary optical excitation experiments on a Mn$_{2}$Au films, N\'{e}el vector aligned in a 60 T magnetic field. In this case laser-irradiation was performed in the absence of an external strain, and XMLD-PEEM images for several marked areas were recorded before and after laser irradiation. Considering the existence of the growth-induced strain, one may expect a similar process of light-assisted DW depinning, resulting in a state with $P \approx 0.4$, as in as-grown films. To have a maximum contrast between $P$ before and after irradiation we investigated Mn$_{2}$Au film, magnetically aligned along the $[110]$ direction in a 60 T magnetic field, with $P \approx 0.8$. 

\begin{figure}[tph]
\centerline{\includegraphics[width=170mm]{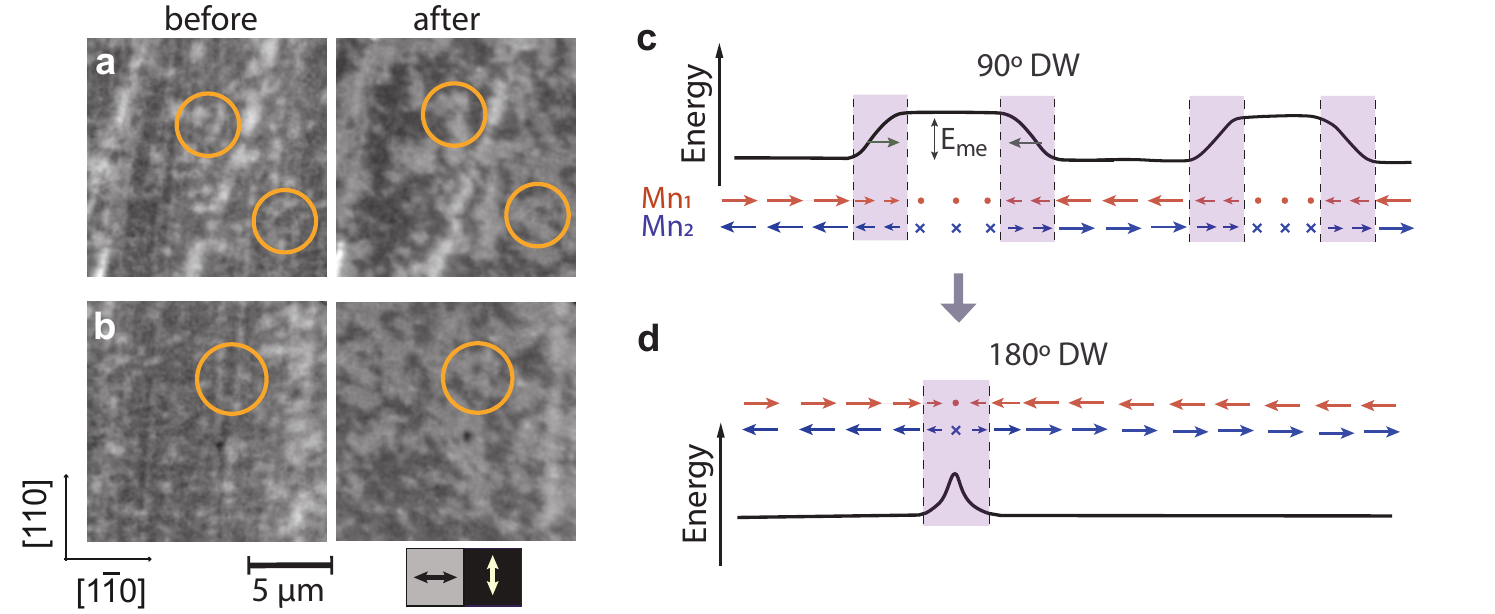}}
\caption{\textbf{Domain wall dynamics following irradiation with optical
pulses.} \textbf{a},\textbf{b} XMLD-PEEM images of a Mn$_{2}$Au film N\'{e}el vector aligned in 60 T magnetic field along the $[110]$ direction recorded before (left) and after (right) irradiation with near-infrared pulse trains with $F=11.5$ mJ/cm$^{2}$ in the absence of external strain. The highlighted regions suggest a growth of the domains aligned along the [1$\overline{1}$0] direction, the direction preferred by the growth-induced strain (see also Fig. \ref{Figure2}). \textbf{c}-\textbf{d} Schematic of the variation of the free energy density under tensile strain, showing also the direction of the motion of 90$^{\mathrm{o}}$ DWs upon light-induced DW depinning (\textbf{c}). The resulting DW motion can result in DW annihilation or formation of 180$^{\mathrm{o}}$ DWs (\textbf{d}), whose position is unaffected by strain.}
\label{Figure3}
\end{figure}

Analyzing XMLD-PEEM images before and after laser irradiation reveals a switching threshold of $\approx 11.5$ mJ/cm$^{2}$. Figure \ref{Figure3} a,b presents the XMLD-PEEM images recorded before and after irradiation for near-threshold excitation at $F = 11.5$ mJ/cm$^{2}$. XMLD-PEEM images of the B-field aligned sample before irradiation (left panels) show predominantly domains with $\textbf{L}
\parallel \lbrack 110]$ ($P\approx 0.8$). After irradiation with $F = 11.5$ mJ/cm$^{2}$ (right panels), the majority of domains have $\textbf{L} \parallel \lbrack 1\overline{1}0]$, with $P\approx 0.4$. Surprisingly, for $F=14$ mJ/cm$^{2}$ the value of $P\approx 0.2$ is reached (see Supplementary Figure \ref{SuppPEEM}), similar to the case of the sample irradiated with $F>5$ mJ/cm$^{2}$ under the applied external strain along the $[110]$ direction. The observed light-induced N\'{e}el vector alignment in the absence of applied strain lends further support to the existence of a weak growth-induced strain in films grown on $r$-cut Al$_{2}$O$_{3}$ substrates. Its presence is likely responsible for the observed light-induced switching of \textbf{L} in the absence of applied strain as well as for the partial alignment of the N\'{e}el vector in as-grown films. Furthermore, the results imply the as-grown state with $P\approx 0.4$ is actually a metastable state, the state with $\textbf{L} \parallel \lbrack 1\overline{1}0]$ being the state with the lowest free energy.

To unambiguously show that it is the photoexcitation-assisted depinning of the DWs that governs the N\'{e}el vector alignment in Mn$_2$Au under strain, an \textit{in situ} time-dependent XMLD-PEEM study would be required. There are, however, several observations that, indirectly, provide support to the proposed mechanism. 

The XMLD-PEEM images of a B-field aligned sample before and after laser irradiation at the threshold fluence (Fig.~\ref{Figure3} a,b) are consistent with the laser-induced domain growth. This can be inferred from areas, highlighted by orange circles, displaying growth of the minority $\textbf{L} \parallel \lbrack 1\overline{1}0]$ domains following photoexcitation of $\textbf{L} \parallel \lbrack 110]$-aligned sample.

Secondly, our proposed mechanism suggests that depinning of the 90$^{\mathrm{o}}$ DWs results in the DW motion in the direction of the free energy gradient (Fig.~\ref{Figure3} c), and subsequent DW annihilation or formation of stable 180$^{\mathrm{o}}$ DWs (Fig.~\ref{Figure3} d). The resulting state should ideally consist of domains with antiparallel N\'{e}el vectors, separated by the 180$^{\mathrm{o}}$ DWs. Indeed, in both current-driven\cite{BodnarCurrentImag} and strain-driven switching experiments worm-like structures are observed in XMLD-PEEM images, attributed to 180$^{\mathrm{o}}$ DWs. A clear support to this assignment is provided by Supplementary Fig. \ref{XMCD}, where the domain structure of the optically switched area is imaged by both XMLD and X-ray magnetic circular dichroism (XMCD) PEEM (see Supplementary Information for details).

Finally, the observation of the switching threshold and its dependence on strain further supports the proposed mechanism. The fact that the magnetically aligned state is clearly observed only for fluences above the threshold fluence (or the related threshold temperature) suggests a thermally activated depinning transition.\cite{Brazovskii} At a fixed temperature, the depinning transition from a pinned DW is realized by increasing the driving force, $G$ (in our case given by the gradient in the magneto-elastic energy, which is proportional to strain), above its threshold value, $G_{T}$, resulting in DW sliding (see Supplementary Fig.~\ref{Depinning}a). It has been demonstrated both theoretically\cite{Ioffe,Vinokur,Brazovskii} and experimentally\cite{Giamarchi,Jeudy} that $G_{T}$ decreases with increasing temperature. At high temperatures $G_{T} \varpropto T^{-\alpha}$ with the exponent $\alpha$ depending on the details of the pinning centers and interactions.\cite{Ioffe,Vinokur,Brazovskii} Let us assume that the application of a tensile strain results in an effective force $G^{*}$, which is lower than $G_{T}$ at the base temperature. This is supported by the lack of permanent magnetic alignment when using only tensile-strain.\cite{Leha_PSS} Photoexcitation results in an increase in local sample temperature, causing a transient reduction in $G_{T}$ to below $G^{*}$ and, thus, launching a sliding DW motion (see Supplementary Fig. \ref{Depinning}). This scenario naturally accounts for the variation in threshold fluence on the applied tensile strain. As $G^{*} \varpropto \varepsilon $ and, thus, differs in the two strain configurations (the growth-induced strain has to be added/subtracted from the applied strain), the observed difference in threshold fluences (temperatures) can be attributed to the presence of the growth-induced strain. In fact, the observation of even higher threshold fluence for switching the B-field aligned film in the absence of external strain further supports this scenario.

\section*{Conclusions}

We present a new approach to locally manipulate the N\'{e}el vector alignment in the collinear metallic antiferromagnet Mn$_{2}$Au by combining strain and excitation with femtosecond laser pulses. To achieve switching, we applied an external strain, which was large enough to induce an additional two-fold anisotropy, yet not strong enough to induce a depinning transition at room temperature. By irradiation of the sample with the train of femtosecond laser pulses we managed to locally align the N\'{e}el vector in two orthogonal orientations, with the direction controlled by the direction of the applied tensile strain, i.e., we realize stable "zero" and "one" states on the same film. This approach allows selective/local control of the N\'{e}el vector axis, with spatial resolution that could be further reduced to a sub-wavelength range by using near-field approaches.

The proposed mechanism of magnetic alignment via DW depinning may also be at play in the current-switching experiments in Mn$_2$Au and CuMnAs.\cite{Wadley_Sci_2016,OleTHz,Stas_Nature,meinert2018,BodnarCurrentImag} While for current-switching of Mn$_2$Au and CuMnAs the driving force is provided by the NSOT, thermal activation could be relevant.\cite{Stas_Nature,meinert2018} In both configurations, however, increasing the effective driving force $G^{*}$, or decreasing of the DW depinning threshold, $G_{T}$, may substantially reduce the required heating to reach domain-wall depinning regime. To this end, further optimization of the film growth, including using different substrate/buffer layer combinations, is required. 

While we demonstrate the N\'{e}el vector alignment triggered by femtosecond laser pulses, and present evidence, suggesting the process is governed by the depinning of domain walls, the details of the underlying process, including timescales, remain to be investigated. To confirm the proposed scenario by recording the dynamics of the domain growth, an ultrafast laser amplifier with sufficient fluence should be coupled to the XMLD-PEEM end-station to enable \textit{in situ} imaging following single laser-pulse excitation. With domains on the micrometer lengthscale and a substantial magnetic linear dichroism in the near-infrared\cite{dichroism} imaging the domain structure with femtosecond near-infrared pulses can also be realized. In this case, combining the optical pump-probe approach with fast modulation of strain (synchronized to the laser repetition rate) could present a way to study domain-wall dynamics on the picosecond timescale. This would be particularly interesting in systems like Mn$_2$Au, where theoretical studies suggest that domain-wall velocities as high as 30 km/s could be achieved.\cite{Gomonay} 

Should such high DW velocities\cite{Gomonay} indeed be realized, a characteristic timescale for optically-induced alignment of micrometer areas would be on the 10-100 ps range, which could enable optical switching with single pulses. Given the fact that the resulting state is stable at room temperature and insensitive to magnetic fields up to several tens of Tesla,\cite{Leha_PSS,Leha_PRB,dichroism} a combination of global strain modulation and local ultrafast manipulation of N\'{e}el vector may provide ways to realize robust high-density memory device with switching timescales in the picosecond range.

\section*{Methods}

\subsection*{Mn$_{2}$Au thin films}

The $c$-axis epitaxial Mn$_{2}$Au thin films are grown on $r$-cut (1$\overline{1}
$02) Al$_{2}$O$_{3}$ substrate, with the lateral size of 10$\times$10 mm$%
^{2} $ and thickness of 530 $\mu$m by the radio-frequency magnetron
sputtering at 600 $^{\circ}$C - see \cite{FilmsAFM} for details. A 40 nm
thick Mn$_{2} $Au film was deposited on a 13 nm thick (001) Ta buffer layer.
To protect the surface, a 2 nm Al layer was deposited on Mn$_{2}$Au, forming
an aluminum-oxide capping layer. Mn$_{2}$Au grows epitaxially with [110] and
[1$\overline{1}$0] axes parallel to the substrate edges, which are along the 
$[211]_{s}$ and $[010]_{s}$ directions of the $r$-cut Al$_{2}$O$_{3}$
substrate, respectively. The $c$-axis of Mn$_{2}$Au films grown on $r$-cut Al$%
_{2}$O$_{3}$ are tilted by 2-3$^{\circ}$ towards the $[010]_{s}$ direction.

\subsection*{Laser irradiation under tensile strain}

The experimental approach to manipulate the N\'{e}el vector by combination of tensile strain and laser pulse excitation is sketched in Fig. \ref{Figure1} a,b.

In the first step, a set of 30 nm thick Cr/Au markers is fabricated on the top of the film using electron beam lithography - see the schematic layout in Fig.~\ref{Figure1}b. These markers allow us to identify distinct areas of approximately $100\times 100$~$\mu $m$^{2}$ in the following steps. The sample is then mounted on a stress device, as used in Ref.~\citenum{Leha_PSS}. Bending the substrate results in tensile strain in the thin Mn$_{2}$Au film along the [110] direction (see Fig.~\ref{Figure1}a). The resulting strain is approximately $\varepsilon =6\pm 1\times 10^{-4}$, as measured by a Si strain gauge attached to the substrate.\cite{Leha_PSS} Then, several marked areas of the film are irradiated by a pulsed laser beam. We use 60 fs pulses at 800 nm and 250 kHz repetition rate. The excitation fluence used in different marked areas is varied between $\approx 3 $ and $\approx 11$ mJ/cm$^{2}$. The linearly polarized laser beam (the effect is found to be independent of the light polarization) is focused onto a spot with $\approx 23$ ~$\mu $m in diameter (full-width at half maximum). Each marked area is irradiated for several seconds by scanning the beam around the center of the marked area, resulting in the size of the irradiated area of about $60-70$~$\mu $m in diameter, as indicated by the colored areas in Fig.~\ref{Figure1}b.

In the next step, the sample is remounted on the strain device and a similar tensile strain is applied along the [1$\overline{1}$0] direction and a different set of marked areas is irradiated (Fig.~\ref{Figure1}a). Finally, a series of marked areas is irradiated without the applied tensile strain. In total, five marked areas on the film are irradiated without applying external strain, five with the tensile strain along the $[110]$ direction and five with tensile strain along the $[1\overline{1}0]$ direction.

A commercial 250 kHz Ti:Sapphire amplifier producing 60 fs laser pulses at $%
\lambda$ = 800 nm (photon energy of 1.55 eV) was used to excite the strained
thin films.

\subsection*{Imaging of Mn$_{2}$Au films using XMLD-PEEM}

The XMLD-PEEM experiments were performed at the SIM beamline of the Swiss
Light Source. The sample was illuminated by linearly polarized X-rays with
both s- and p- polarizations. The angle of incidence was 16$^{\circ}$. To
achieve the best contrast, we utilized a slightly modified method than that
described in Ref.~\citenum{BodnarCurrentImag}. Namely, we acquired images with
two orthogonal linear polarizations of the X-rays at each of the two photon
energies (in this case 637.6 eV and 638.6 eV) around the $L_{3}$ absorption
edge of Mn, corresponding to maximal XMLD contrast.\cite{Leha_PSS} The
resulting XMLD-PEEM images were obtained by pixelwise division of XMLD contrast maps recorded at 638.6 eV and 637.6 eV, where each XMLD was determined by the normalized difference of images acquired with the two orthogonal X-ray polarizations. The field of view used is 20 $\mu$m.

\bibliographystyle{plain}
\bibliography{publist}

\section*{Acknowledgments}

This work was funded by the Deutsche Forschungsgemeinschaft (DFG, German
Research Foundation) Grant No. TRR 173 - 268565370 (project A05, with
contribution from A01) and TRR 288 - 422213477 (project B08) and received support from Horizon 2020 Framework Program of the European Commission under grant agreement No. 863155 (S-NEBULA). V.G. and M.F. acknowledge the financial support from the
Graduate School of Excellence "Materials Science in Mainz" (DFG GSC 266
49741853). P.G. gratefully acknowledge support from the Agence Nationale de Recherche, via the MeMoPas project ANR-19-CE46-0006-1, as well as access to the HPC resources of IDRIS under the allocation A0090910965 attributed by GENCI. We acknowledge the Paul Scherrer Institute, Villigen, Switzerland for the beamtime allocation under proposal 20200977 at the SIM beamline of the SLS. The authors thank the SIM beamline staff for the technical support. We acknowledge the support of the HLD at HZDR, member of the European Magnetic Field Laboratory (EMFL).
We acknowledge valuable discussions with H. Gomonay and D. Fuchs.

\section*{Author contributions}

V.G. and J.D. conceived the project. V.G., M.F., Y.L., S.S., and A.K. performed XMLD-PEEM measurements. Y.L. and S.B. grew the samples and made masks. V.G. and A.R.P. performed laser irradiation experiments under strain. Y.S. aligned the samples in high magnetic field. V.G. and M.F. analyzed the XMLD-PEEM data with input from M.J.. P.G. performed DFT calculations. A.S. and J.D. performed laser heating simulations. V.G. and J.D. wrote the manuscript with contributions from all coauthors.

\section*{Competing interests}

The authors declare no competing interests.

\section*{Data availability}

All relevant data are available from the authors.

\newpage

\section*{Supplementary Information}
\renewcommand{\figurename}{Supplementary Figure}
\setcounter{figure}{0}

\subsection*{Free energy landscape in a strained easy-plane AFM}

The relevant components of the free energy functional are the
magneto-crystalline anisotropy energy, $E_{anis}$, and the magneto-elastic
energy, $E_{me}$, and could be written as follows: 
\begin{equation}
E=E_{anis}+E_{me},
\end{equation}%
For a tetragonal material, the anisotropy energy is given by:

\begin{equation}
E_{anis}=K_{2\bot }cos^{2}(\Theta )+K_{4\bot }cos^{4}(\Theta
)+K_{4||}sin^{4}(\Theta )cos(4\phi ),
\end{equation}%
where $\Theta $ and $\phi $ are the axial and polar angles of the
corresponding spin, $K_{2\bot }$ and $K_{4\bot }$ are out-of-plane and $%
K_{4||}$ in-plane magneto-crystalline anisotropy constants. For the case of
an easy plane AFM, $E_{anis}$ reduces to: 
\begin{equation}
E_{anis}=K_{4||}sin^{4}(\Theta )cos(4\phi ).
\end{equation}%
For Mn$_{2}$Au, $K_{4||}$ was measured to be $1.8$~$\mu $eV per formula unit.%
\cite{Sempa}

The magneto-elastic energy $E_{me}$ is given by: 
\begin{equation}
E_{me}=B_{ij}\epsilon _{ij}\beta _{i}\beta _{j},
\end{equation}%
where $\beta _{ij}$ are the directional cosines. In case of unidirectional
strain, the strain tensor has only one component, $\epsilon _{xx}$. Thus,

\begin{equation}
E_{me}=B_{me}\epsilon _{xx}cos^{2}(\phi ).
\end{equation}

In the absence of strain, the free energy exhibits a four-fold symmetry,
which should result in equal distribution of the [110] and [1$\overline{1}$%
0] domains. Under the external unidirectional strain, the free energy
landscape is modified such, that the local minima for the [110] and [1$%
\overline{1}$0] domains become non-degenerate. Using DFT calculations, we
estimate $B_{me}=0.9$~meV per formula unit, resulting, in $B_{me}\epsilon
_{xx}=0.6 \mu $eV per formula unit for $\epsilon _{xx}=6\times 10^{-4}$
which is comparable to $K_{4||}$. Moreover, given the weak in-plane
anisotropy in Mn$_{2}$Au, even a weak growth-induced strain can lead to an
imbalance between the volume fractions of the two types of domains, as
experimentally observed.

\subsection*{Density Functional Theory Calculations}

Vienna ab-initio simulation package (VASP) \cite{Kresse} was used to
describe electron exchange and correlations within the
Perdew-Burke-Ernzerhof (PBE) generalized gradient approximation, together
with Projector Augmented Wave (PAW) \cite{Perdew} basis set with the cut-off
energy at 600 eV. The Brillouin zone was sampled with $25\times 25\times 25$
Monkhorst-Pack k-point grid. The values of these parameters were chosen
following a series of convergence tests on forces with a tolerance of a few
meV/\textup{\AA }. Firstly, the equilibrium unit cell parameters where
determined by minimizing the strain with respect to the unit cell
parameters. At this stage the occupations were smeared with a
Methfessel-Paxton scheme of the order one with a smearing width of 0.1 eV.
The minimized unit cell is given by the centered tetragonal structure (bct$%
_{2}$) with lattice parameters a = 3.27929\textup{~\AA } and c = 8.43087%
\textup{~\AA }. The cell has zero total magnetization, with Au atoms
carrying zero moment and plus/minus 3.55 $\mu $B for Mn atoms. The atomic
positions were then relaxed with the tolerance on maximum force of 1 meV/%
\textup{\AA }. The elastic constants were calculated by fitting the
strain-strain dependence for a number of strained configurations giving the
following values: C$_{11}$ = 138 GPa, C$_{12}$ = 126 GPa, C$_{13}$ = 75 GPa,
C$_{33}$ = 232 GPa, C$_{44}$ = 82 GPa, and C$_{66}$ = 110 GPa.

\begin{figure}[tph]
\centerline{\includegraphics[width=100mm]{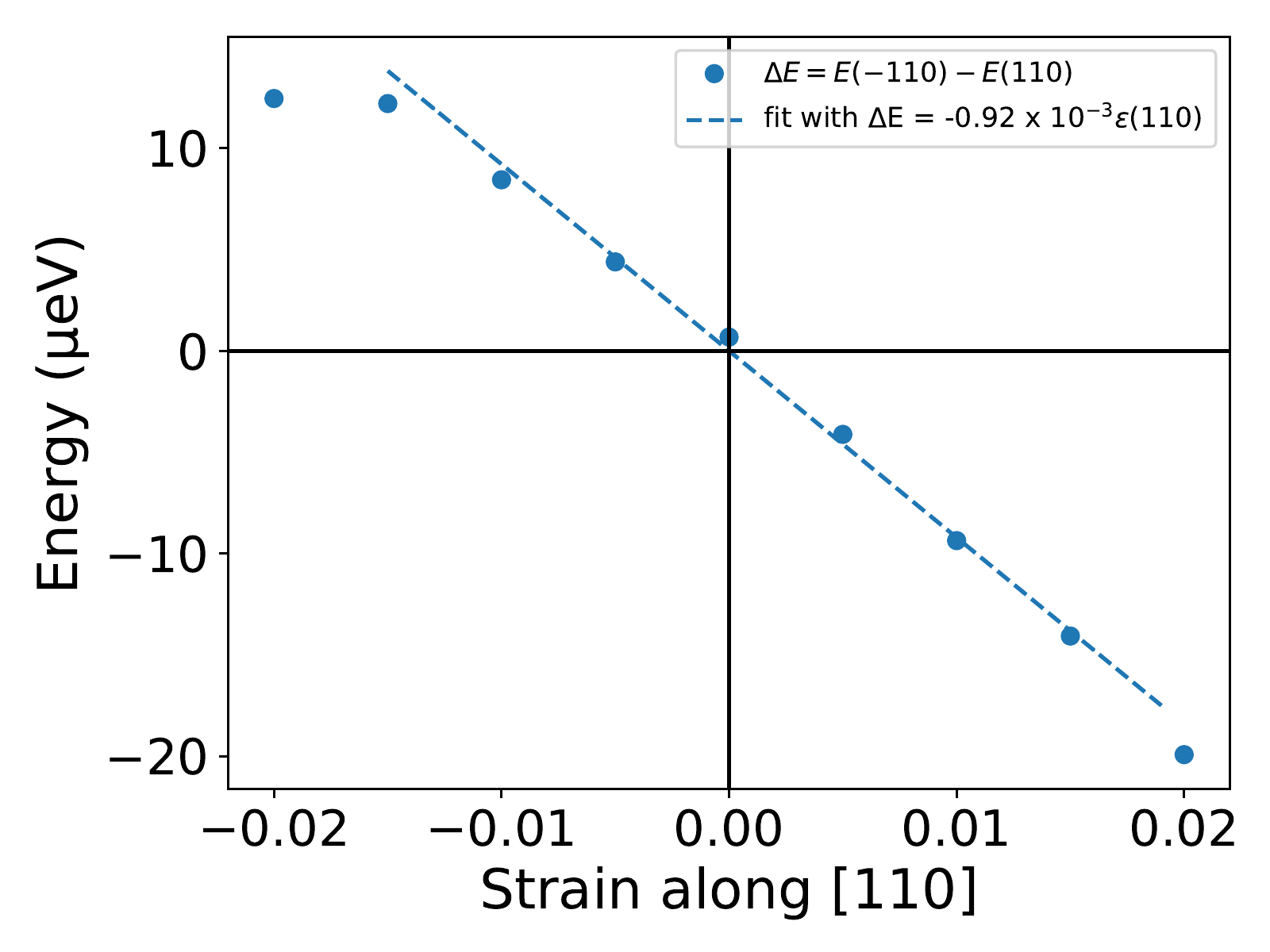}}
\caption{\textbf{DFT calculation of the magneto-elastic coefficient.} }
\label{Magnetoelastic}
\end{figure}

To calculate the magnetic anisotropy spin-orbit coupling is taken into
account. Here, the tetrahedron method with Bl\"{o}chl corrections for
smearing was used instead of the Methfessel-Paxton scheme. Electronic energy
minimization convergence criterion was set to 10$^{-9}$ eV. At this stage
the energy convergence of a few $\mu $eV with respect to k-points and energy
cut-off was achieved. The energy difference for the configurations with
spins oriented along [001] and along [110] directions is found to be 2.55
meV, in excellent agreement with reported values.\cite{Shick}

The energy difference between the configuration with magnetic moments
oriented along the [1$\overline{1}$0] direction E(1$\overline{1}$0) and
along [110] direction E(110) were obtained for a number of cells strained
along the [110] direction. Here, a positive value of $\mathrm{\Delta E=E(1%
\overline{1}0)-E(110)}$ indicates that the configuration with magnetic
moments oriented along the [110] direction are stable while for negative
values configurations with [1$\overline{1}$0] are energetically favorable -
see Supplementary Fig.~\ref{Magnetoelastic}. The linear fit results in $-9.2 \mu $eV per
one percent of magnetic anisotropy with strain applied along $\langle ${110}$%
\rangle $ directions.

Magnetostriction coefficient $\mathrm{\lambda ^{\sigma ,2}=-23.38\times
10^{-6}}$ as well as the magneto-elastic coefficient $b_{3}$ = 5.1 MPa were
also obtained, with the procedure described in detail in Ref. \citenum{Nieves}.

\subsection*{XMLD-PEEM data on laser assisted Néel vector switching of B-field aligned samples}

As described in the main text, the threshold for switching the Néel axis from [110] to $[1\overline{1}0]$ in the absence of applied strain was found at $F = 11.5$ mJ/cm$^2$ (see Fig. \ref{Figure3} of the main text). Above threshold, the area of domains aligned parallel to $[1\overline{1}0]$ axis further increases, reaching values obtained in the sample aligned under the applied strain along [110] directions (Néel vector aligns perpendicular to the applied tensile strain). The corresponding XMLD-PEEM images of the are before and after irradiation are shown in Supplementary Fig. \ref{SuppPEEM}.

\begin{figure}[ht]
\centerline{\includegraphics[width=110mm]{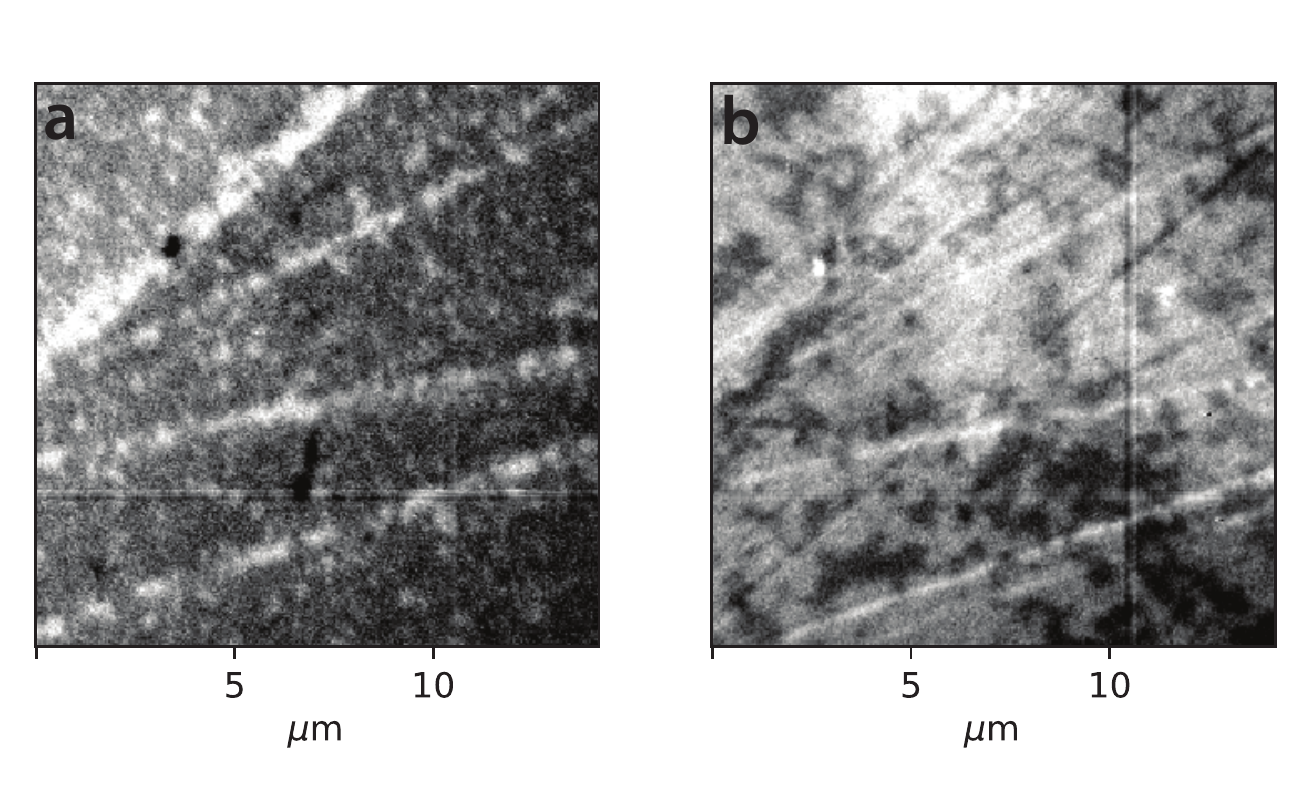}}
\caption{XMLD-PEEM images of a Mn$_{2}$Au film N\'{e}el vector aligned
in 60 T magnetic field along the [110] direction recorded before (\textbf{a}) and after (\textbf{b}) irradiation by the near-infrared pulses with excitation fluence of 14 mJ/cm$^{2}$ in the absence of external strain. Compared to the data with excitation fluence of 11.5 mJ/cm$^{2}$ (Fig. \ref{Figure3} in the main text) excitation results in nearly fully polarized state.}
\label{SuppPEEM}
\end{figure}

\subsection*{Domain structure of laser-aligned sample probed by X-ray magnetic circular dichroism (XMCD)}

In addition to the XMLD contrast we recorded X-ray magnetic circular dichroism (XMCD) contrast images of Mn$_{2}$Au magnetically aligned by laser irradiation under strain. Comparison of the XMCD contrast images with the XMLD contrast images of areas, presented in Supplementary Fig. \ref{XMCD}, reveals that all narrow worm-like domains seen in XMLD-PEEM correspond to the 180$^{\mathrm{o}}$ domain walls between the two anti-parallel oriented domains.

\begin{figure}[h]
\centerline{\includegraphics[width=110mm]{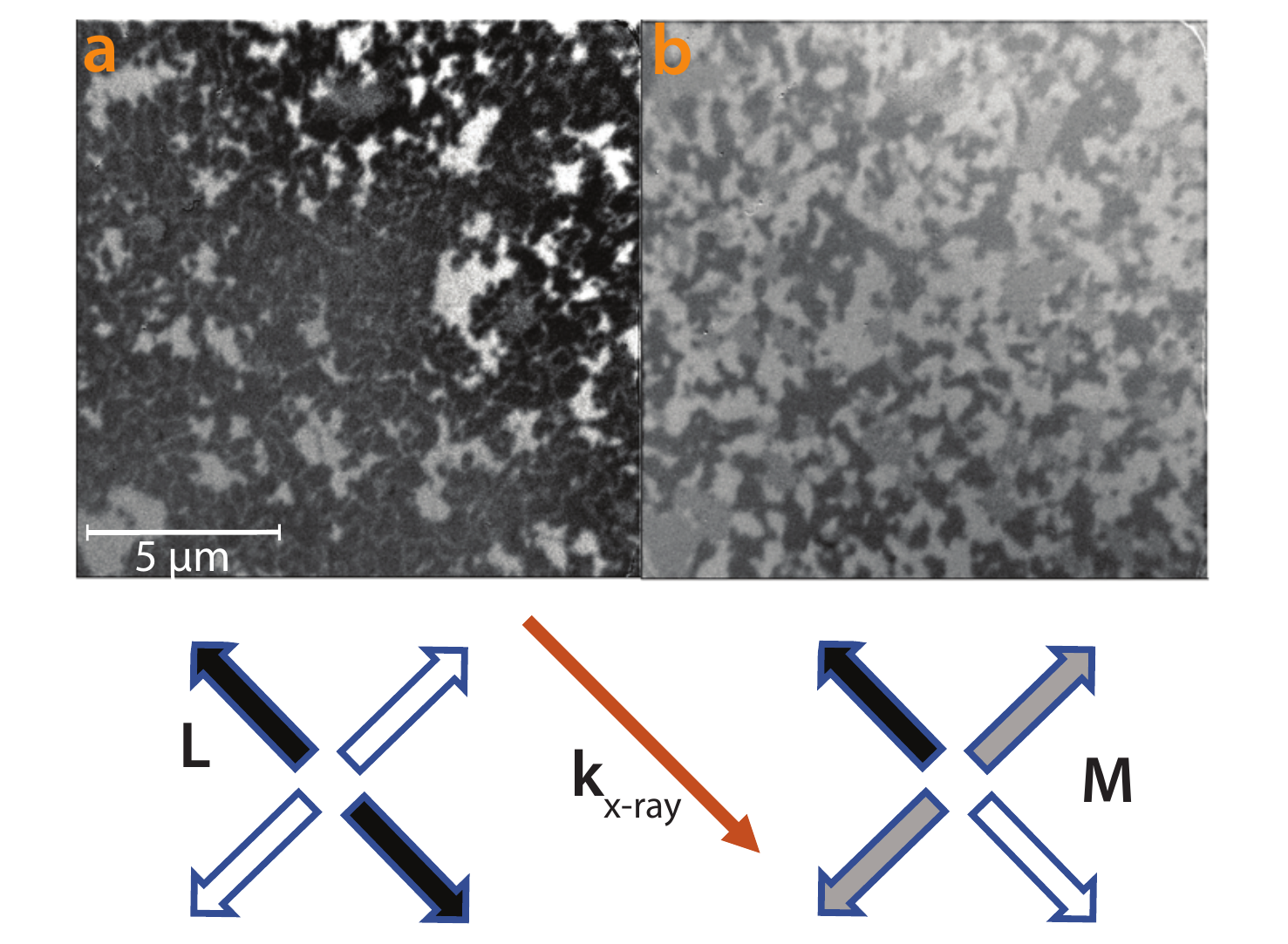}}
\caption{\textbf{Comparison of the XMLD-PEEM (left) and XMCD-PEEM (right) images recorded on laser/strain aligned Mn$_{2}$Au.} The color coding represents the orientation of the N\'{e}el vector (\textbf{L}) and the uncompensated surface magnetization (\textbf{M}) relative to the X-ray propagation direction.}
\label{XMCD}
\end{figure}

As shown in Supplementary Fig. \ref{XMCD}b, the XMCD contrast gives three distinct color levels in the PEEM image. In this XMCD configuration (see the schematic below the XMCD-PEEM image) the black and the white areas correspond to domains with N\'{e}el vectors laying in the plane of incidence of the X-ray beam, aligned anti-parallel to each other. These two domain types are indistinguishable in the XMLD image (Supplementary Fig. \ref{XMCD}a). Gray areas in the XMCD-PEEM image correspond to domains with the N\'{e}el vector perpendicular to the X-ray propagation direction. For these domains, the exact direction in which the N\'{e}el vector is pointing cannot be resolved in this experimental geometry. These two domain types can be, however, distinguished by rotating the sample by 90$^{\mathrm{o}}$ (or, alternatively, by changing the propagation direction of the X-ray beam). 

In a systematic study of the observed XMCD effect, which will be published elsewhere, the XMCD contrast could be associated with the uncompensated surface moments in Mn$_2$Au. 

\subsection*{Schematic description of the thermally-assisted depinning transition}

\begin{figure}[tph]
\centerline{\includegraphics[width=140mm]{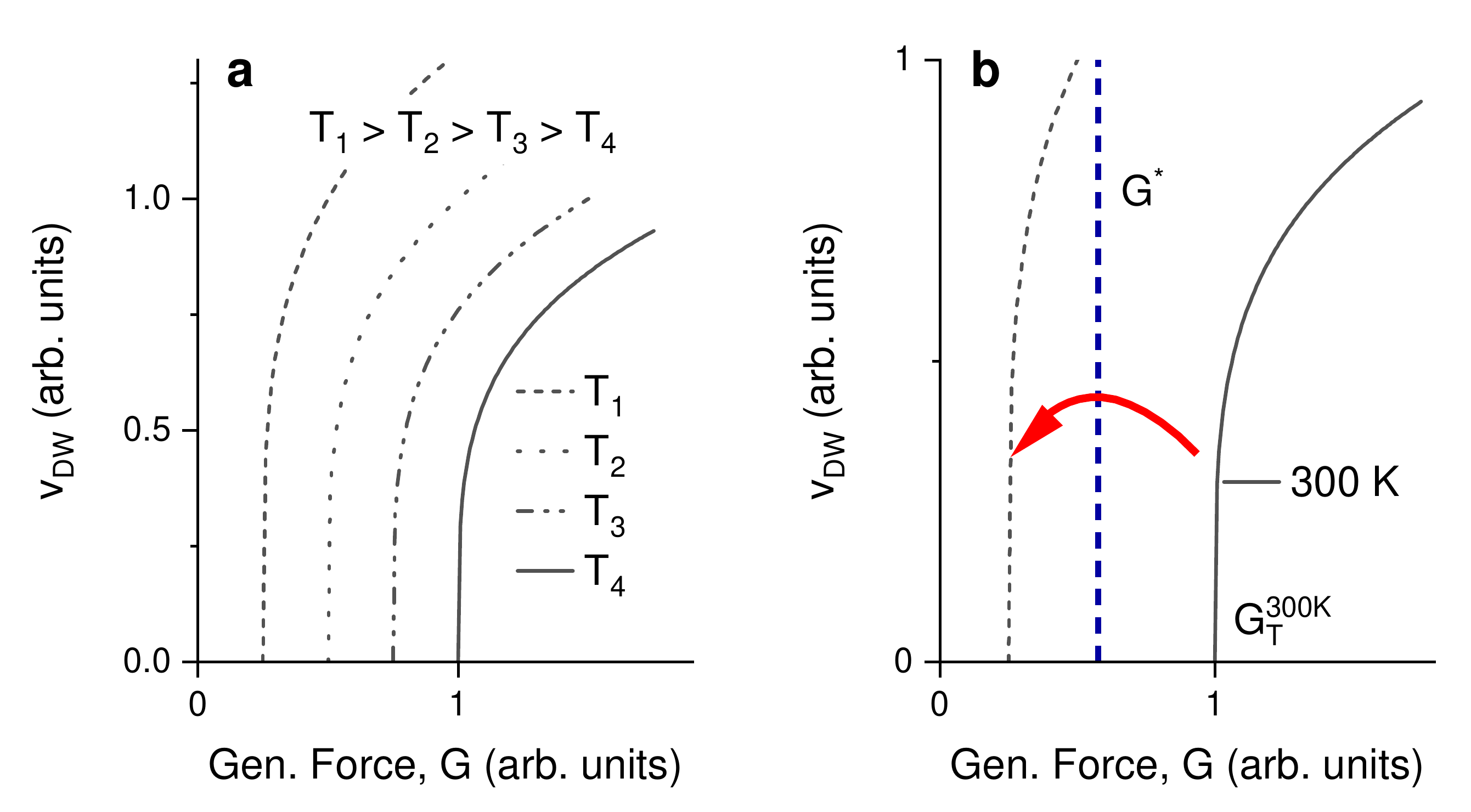}}
\caption{\textbf{Schematic description of the photoassisted depinning transition.} \textbf{a} Schematic dependence of the domain-wall velocity (v$_{DW}$) on the applied generalized force, $G$, for different base temperatures. For simplicity, we assume $v_{DW}=0$ below the threshold force, $G_{T}$, i.e., we neglect the sub-threshold thermally activated creep motion.\cite{Jeudy} In the case of tensile strain, $G$ is provided by the strain-induced gradient of the free energy density. \textbf{b} At room temperature, the applied strain results in a generalized force $G^{*}$ (blue dashed vertical line), which is lower than the threshold depinning force at 300 K, $G_{T}^{300K}$. Upon photoexcitation, the resulting temperature increase shifts the threshold force $G_{T}$ to below $G^{*}$, resulting in a depinning transition and growth of energetically favorable domains.}
\label{Depinning}
\end{figure}

Here we sketch the proposed mechanism of thermally induced DW depinning transition. The depinning transition from a pinned DW (where the DW velocity $v_{DW}=0$) to a sliding DW is realized by increasing the driving force above its threshold value, $G_{T}$. Supplementary Fig.~\ref{Depinning}a presents the schematic dependence of $v_{DW}(G)$ at different temperatures.\cite{Brazovskii,Ioffe,Vinokur,Giamarchi,Jeudy} For simplicity, we neglect the sub-threshold thermally activated creep motion.\cite{Jeudy} Supplementary Fig.~\ref{Depinning}b sketches the effect of optical excitation on DW motion at room temperature, under the applied external force $G^{*}<G_{T}^{300K}$. Here, the resulting transient heating reduces $G_{T}$ to a value lower than $G^{*}$, resulting in DW depinning. Naturally, the threshold fluence (final temperature) to achieve DW depinning depends on $G^{*}$, accounting for a difference in threshold fluences in different strain configurations.

\subsection*{Estimates of laser heating effects}

When irradiating the sample with a femtosecond laser beam there are several
heating effects to be considered. The first is the continuous sample
heating, which is proportional to the average irradiation intensity. The
continuous laser heating in a thin film is governed by the absorbed light
intensity and the thermal properties of the substrate. The temperature
increase of the illuminated region can be estimated using a simple
steady-state heat diffusion model.\cite{heating,mihailovic} Using a
reflectivity of 0.52 and the thermal conductivity, $\kappa $, of sapphire at
room temperature of 35 W/mK \cite{kappa} we estimate the steady state
heating to be $<6$ K over the entire range of fluencies used in this
experiment. Thus, the continuous laser heating effects can be neglected.

To estimate transient laser heating in a metallic sample, two effects can be
considered. On a femtosecond timescale, the rapid electron-electron
thermalization leads to a rapid thermalization of the electron gas. Here,
assuming no heat transfer to the lattice, the resulting electronic
temperature can be estimated considering the absorbed energy per pulse and
the electronic specific heat. Considering the calculated electronic density
of states in Mn$_{2}$Au \cite{Stas_Nature} and the Sommerfeld model for the
specific heat we obtain the resulting electronic temperatures between 3800
and 6700 K for excitation fluence between 4.2 and 12.6 mJ/cm$^{2}$. While
the optical penetration depth of Mn$_{2}$Au is estimated to be approximately 30
nm, we assume that the hot electron transport results in a homogeneous
electronic temperature throughout the film/buffer layer thickness.

On the timescale of a few picoseconds, the electrons thermalize with the
lattice, resulting in a temperature that is governed by the total specific
heat of the metallic layer.\cite{Obergfell} Under these assumptions, the
temperature increase, $\Delta T$, is estimated from the absorbed energy
density:

\begin{equation*}
\Delta T=\frac{F\left( 1-R-T\right) }{C_{p}^{Mn_{2}Au}\ast
d_{Mn_{2}Au}+C_{p}^{Ta}\ast d_{Ta}}.
\end{equation*}

Here, $R\approx 0.52$ is the reflectivity,\cite{dichroism} $T\approx 0.03$
is the transmission, $d_{Mn_{2}Au}$, $d_{Ta}$ are the thicknesses and $%
C_{p}^{Mn_{2}Au}$, $C_{p}^{Ta}$ are the total specific heats of the
corresponding metallic layers. As at temperatures above 300 K the lattice
specific heat dominates, we use the Doulong-Petit limit. The resulting
temperatures as a function of irradiation fluence $F$ are presented in Fig.~\ref{Figure2} b. Note that the subsequent cooling of the excited sample volume is
governed by the heat diffusion into the substrate, governed by the
substrate's thermal conductivity, $\kappa $. To estimate the evolution of
temperature we use a simple heat diffusion model for a semi-infinite solid:%
\cite{mihailovic}

\begin{align}
{\Delta T}(\mathbf{r},t)   &  =\frac{P_{0}\alpha\left(
1-R\right)  }{4\pi\rho_{s}c}e^{-\alpha z}\nonumber\\
&  \times\int\limits_{-\infty}^{t}dt^{\prime}\frac{\exp\left(  -\frac
{2t^{\prime2}}{\tau^{2}}+{\alpha}^{2}{d}_{i}\left(  t-t^{\prime}\right)
-\frac{x^{2}}{4\left(  t-t^{\prime}\right)  d_{i}+d^{2}}-\frac{y^{2}}{4\left(
t-t^{\prime}\right)  d_{i}+d^{2}}\right)  }{\sqrt{{4}\left(  {t-t}^{\prime
}\right)  {d}_{i}{+d}^{2}}+\sqrt{{4}\left(  {t-t}^{\prime}\right)  {d}_{i}%
+{d}^{2}}}\label{dTpulse}\\
&  \times\lbrack1+\operatorname{erf}(\frac{{2z-4\alpha d}_{i}\left(
{t-t}^{\prime}\right)  }{{4}\sqrt{\left(  {t-t}^{\prime}\right)  {d}_{i}}%
})+e^{2\alpha z}(1-\operatorname{erf}(\frac{{2z+4\alpha d}_{i}\left(
{t-t}^{\prime}\right)  }{{4}\sqrt{\left(  {t-t}^{\prime}\right)  {d}_{i}}%
}))]\;,\nonumber
\end{align}

Here $d_{i}$ = $\kappa/\rho_{s}c$ is the thermal diffusivity, $\rho_{s}$ is
the density of sapphire, $\alpha$ is the absorption coefficient and
$\operatorname{erf}(x)=\int\limits_{0}^{x}\exp\left(  -t^{2}\right)  dt$.

\begin{figure}[pth]
\centerline{\includegraphics[width=80mm]{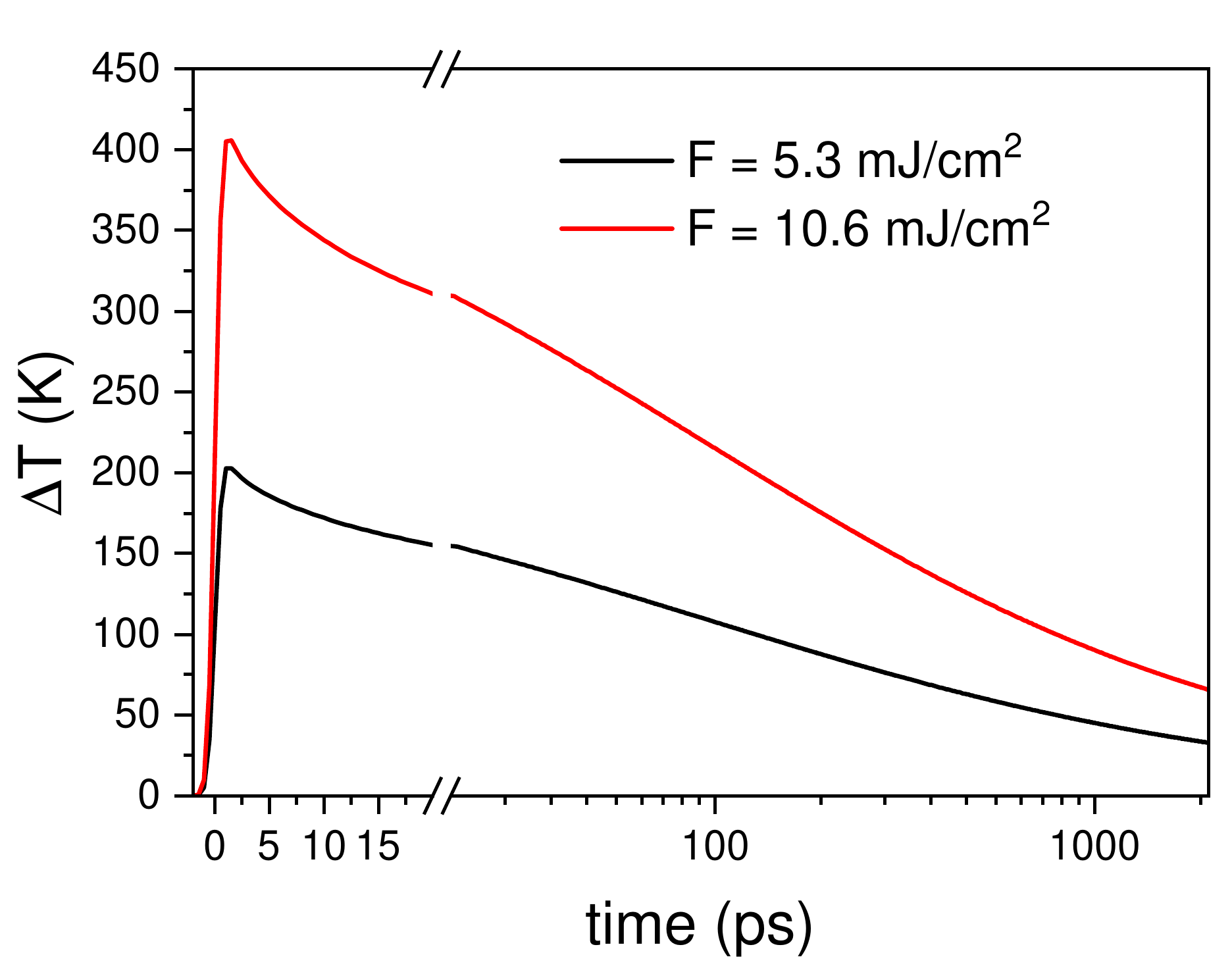}}
\caption{\textbf{Simulated temporal evolution of temperature of the
irradiated spot for two characteristic excitation fluences.} Time evolution
of the temperature change of the film is calculated for two characteristic
excitation fluences using a simple heat diffusion model,\protect\cite%
{mihailovic} given by Eq. \protect\ref{dTpulse}. Note the axis
break at 20 ps, accompanied by the change in scale from linear to
semi-logarithmic.}
\label{Heating}
\end{figure}

It follows from Supplementary Fig.~\ref{Heating} that the characteristic time, at which $%
\Delta T$ drops to one half of its peak value is of the order of 100 ps.
\end{document}